\documentclass[]{spie}  %>>> use for US letter paper
\usepackage{amsmath,amsfonts,amssymb}
\usepackage{graphicx}
\usepackage{tabularx}
\usepackage{siunitx}
\usepackage{array}
\usepackage[colorlinks=true, allcolors=blue]{hyperref}
\usepackage[export]{adjustbox}
\usepackage{makecell}

\title{The Calibration System of the iLocater Spectrograph}

\author[a]{Sai Vidyud Senthil Nathan}
\author[a]{Jonathan Crass}
\author[a]{Julia E. Brady}
\author[a]{Mark Derwent}
\author[a]{Marcelo Tala Pinto}
\author[b]{Brian Sands}
\author[a]{Jackson Datkuliak}
\author[a]{Jonathan Shover}
\author[c]{Christian Schwab}
\author[d]{Julian St\"urmer}
\author[e]{Stanimir O. Letchev}
\author[e]{Jacob Pember}
\author[a, f]{Jayde Spiegel}
\author[a]{Erin Duell}
\author[a]{Daniel Pappalardo}
\author[a]{Marshall C. Johnson}
\author[a]{Michael Engelman}
\author[d, g]{Matheus J. Castro}
\author[h]{Justin R. Crepp}
\author[c]{Ondrej Kitzler}
\author[i]{Thomas Legero}
\author[a, j]{Xavier Lesley-Salda\~na}
\author[a, j]{Richard Pogge}
\author[c]{Dane Zielinski-Nicolson}

\affil[a]{Department of Astronomy, The Ohio State University, 140 West 18$^{\mathrm{th}}$ Ave., Columbus, OH 43210, USA}
\affil[b]{Engineering and Design Core Facility, University of Notre Dame, Notre Dame, IN 46556, USA}
\affil[c]{School of Mathematical and Physical Sciences, Macquarie University, Sydney, Australia}
\affil[d]{Landessternwarte, Zentrum f\"ur Astronomie der Universit\"at Heidelberg, K\"onigstuhl 12, 69117 Heidelberg, Germany}
\affil[e]{Max-Planck-Institute for Astronomy, Königstuhl 17, 69117, Heidelberg, Germany}
\affil[f]{Physics Department, Un Santa Cruz, 1156 High Street, Santa Cruz, CA 95064, USA}
\affil[g]{Fakult\"at f\"ur Physik und Astronomie, Universit\"at Heidelberg, Im Neuenheimer Feld 226, 69120 Heidelberg, Germany}
\affil[h]{Department of Physics \& Astronomy, University of Notre Dame, 225 Nieuwland Science Hall, Notre Dame, IN 46556, USA}
\affil[i]{Physikalisch-Technische Bundesanstalt, Bundesallee 100, 38116 Braunschweig, Germany}
\affil[j]{Center for Cosmology and AstroParticle Physics, The Ohio State University, 191 West Woodruff Avenue, Columbus, OH 43210}

\authorinfo{Further author information: (Send correspondence to S.V.S.N or J.C.) \\ S.V.S.N.: E-mail: saividyud@g.ucla.edu \\ J.C.: E-mail: crass.7@osu.edu}

\begin{document} 
\maketitle

\begin{abstract}

iLocater is a new near-infrared, extreme precision radial velocity (EPRV) instrument delivered to the Large Binocular Telescope in the summer of 2026. The instrument utilizes single-mode fibers (SMFs) for light transmission, including its calibration systems. We present an overview of the iLocater calibration system, detailing the specific sources and the hardware required to inject calibration light into the SMFs. Furthermore, we discuss the hardware that efficiently switches between these sources and distributes the light to all necessary instrument locations, entirely eliminating the need for free-space propagation.

\end{abstract}

\keywords{Infrared instrumentation, EPRV spectrographs, wavelength calibration, single-mode fibers}

%=========================================================================
\section{Introduction} \label{sec:introduction}

The iLocater spectrograph \cite{Crepp2016, Crass2022} is a near-infrared (\qty{0.97}-\qty{1.31}{\micro\metre}), extreme precision radial velocity (EPRV) instrument recently deployed at the Large Binocular Telescope (LBT). The instrument is fed by the LBT's adaptive optics system that couples light into single-mode fibers (SMFs) for spectrograph illumination. Given the small core diameter of these SMFs ($\sim$\qty{6}{\micro\metre}), efficiently injecting light and minimizing throughput losses across different calibration sources pose a significant challenge. To address this, iLocater utilizes a custom fiber switching system to keep light within the SMFs after injection, eliminating the need for additional free-space coupling. Here, we present an overview of the iLocater calibration sources (\S\ref{sec:calibration_sources}), the high-level design and requirements of the fiber switching system (\S\ref{sec:switching_system_schematic}), and the final design and implementation of the system (\S\ref{sec:mechanical_design_and_hardware}). We discuss laboratory testing performance (\S\ref{sec:preliminary_laboratory_testing_results}) and close with conclusions and future work (\S\ref{sec:conclusion_and_future_work}).

%=========================================================================
\section{Calibration Sources} \label{sec:calibration_sources}

The iLocater spectrograph receives light via three input fibers: the left (SX) side of the LBT, the right (DX) side of the LBT, and a simultaneous calibration fiber. To enable EPRV science, each fiber must be precisely calibrated. This is achieved through a daily calibration routine using five distinct sources: a Fabry-Pérot etalon (\S\ref{subsec:etalon}), a halogen light source (\S\ref{subsec:halogen_light_source}), a laser frequency comb (\S\ref{subsec:laser_frequency_comb}), a uranium neon hollow cathode lamp (\S\ref{subsec:uranium_neon_cathode_lamp}), and a solar feed (\S\ref{subsec:solar_feed}).

\begin{figure}[p]
    \centering

    \begin{tabular}{c c} %% tabular useful for creating an array of images 
        \includegraphics[width=0.42\textwidth]{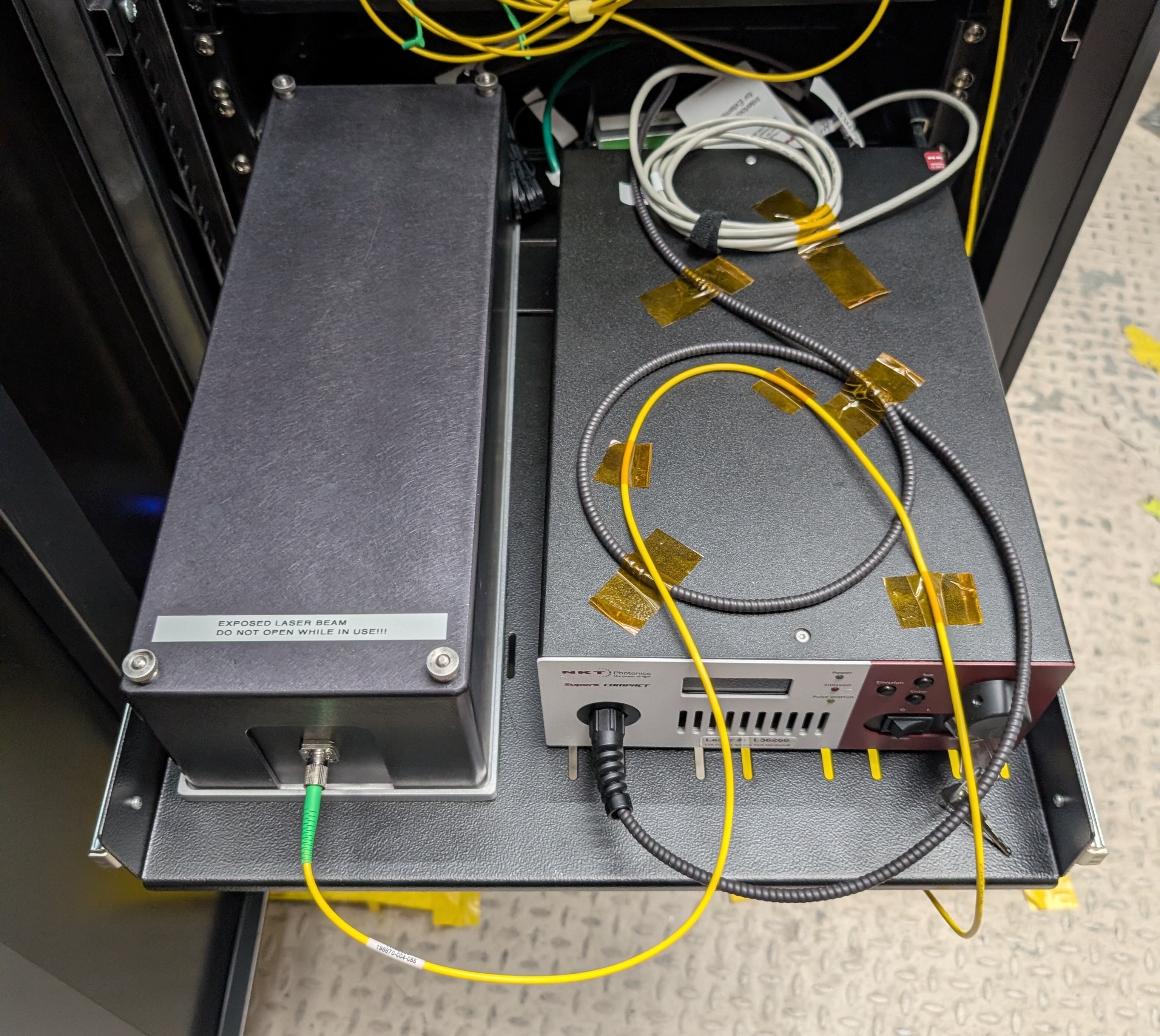} & \includegraphics[width=0.35\textwidth]{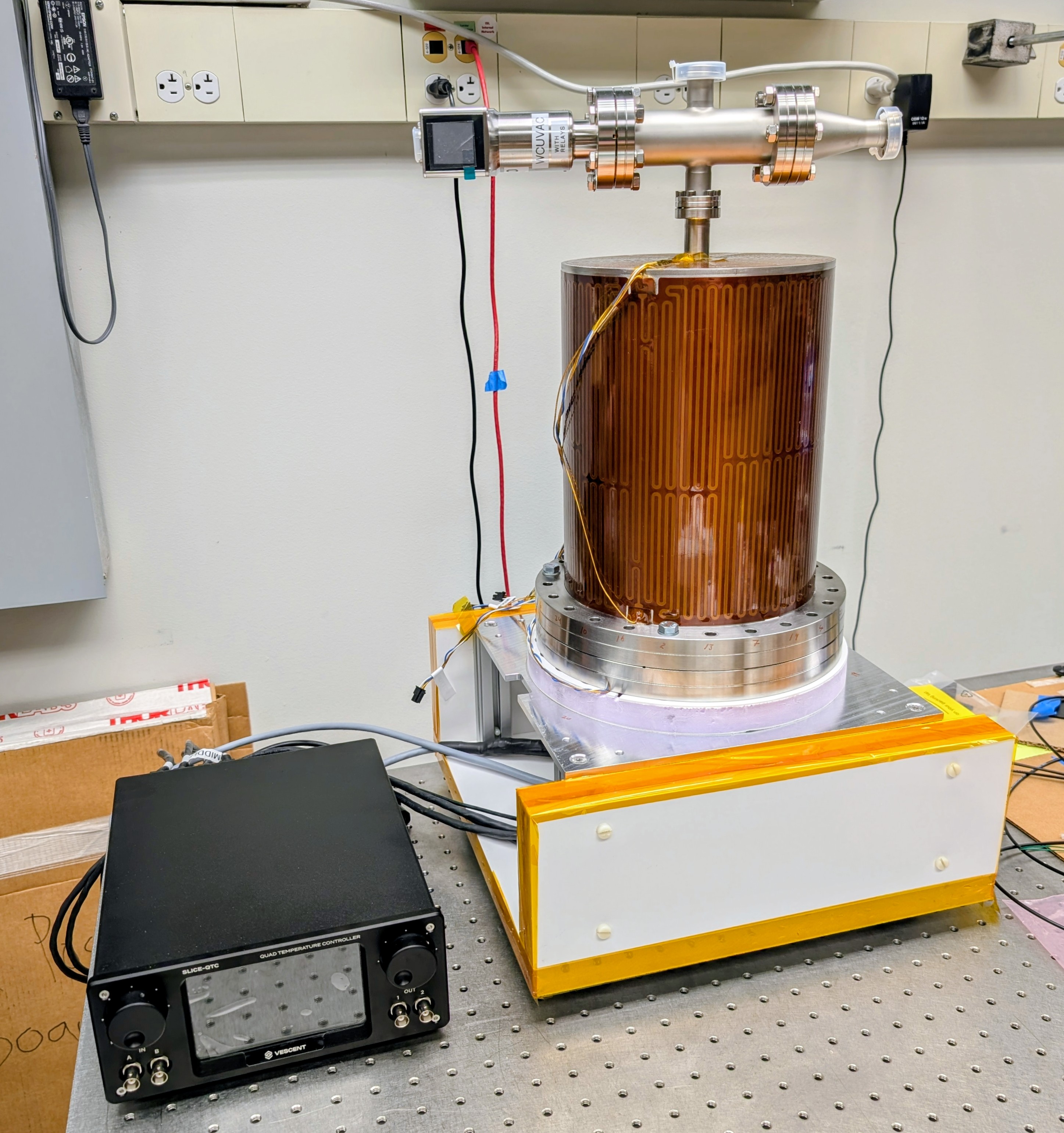}
    \end{tabular}
    \caption{Left: The SuperK COMPACT white light laser and injection optics system (enclosed). Right: The Fabry-Pérot etalon system housed in its thermally stabilized vacuum enclosure.}
    \label{fig:superk_compact_laser_and_etalon}
    
    \vspace{0.6cm} % Adds spacing between the figures

    \begin{minipage}{0.47\textwidth}
        \centering
        \includegraphics[width=\textwidth]{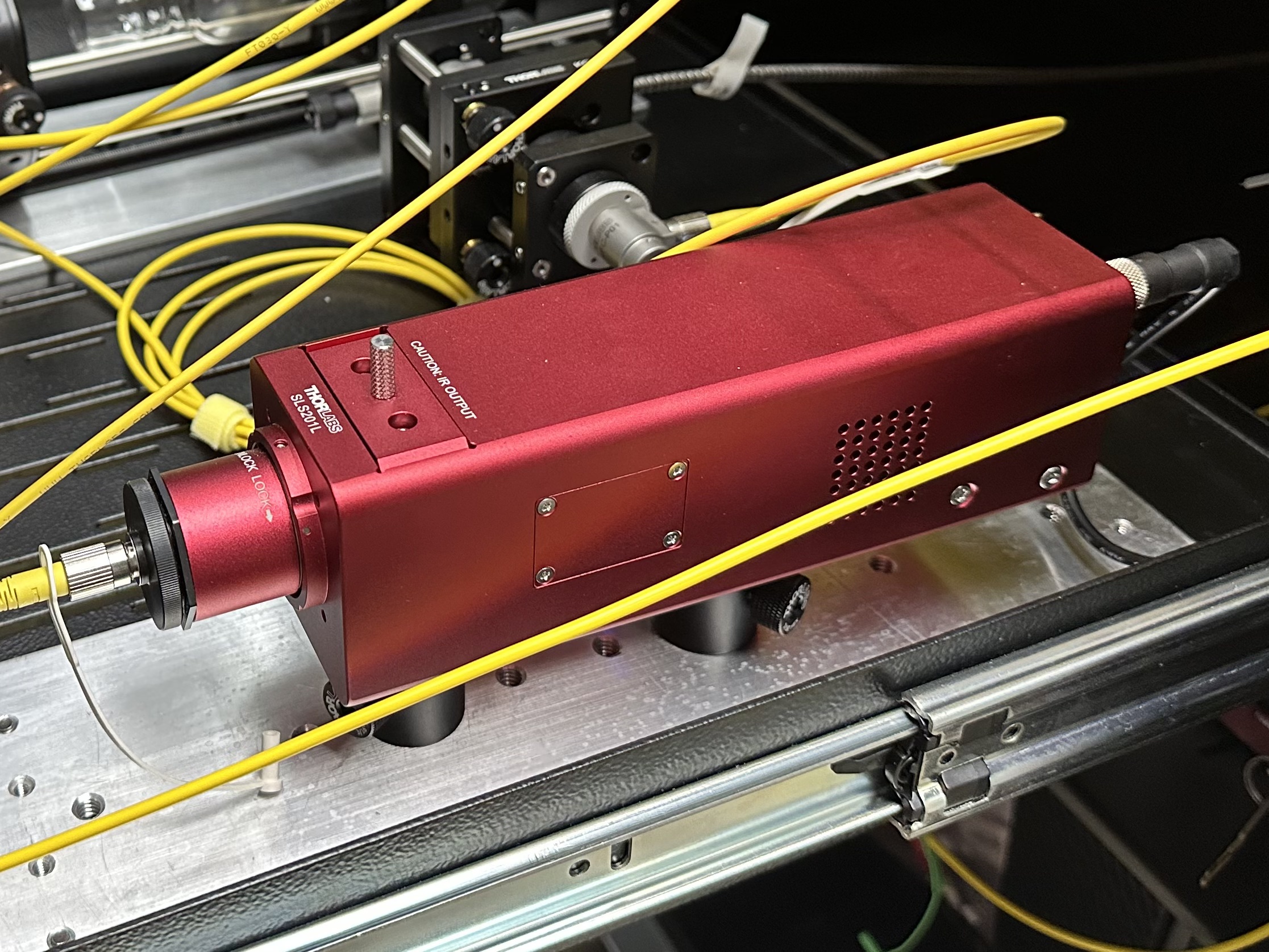}
        \caption{The iLocater tungsten halogen calibration source from Thorlabs.}
        \label{fig:halogen}
    \end{minipage}\hfill % Fills the space between them
    \begin{minipage}{0.47\textwidth}
    \centering
        \includegraphics[width=0.78\textwidth]{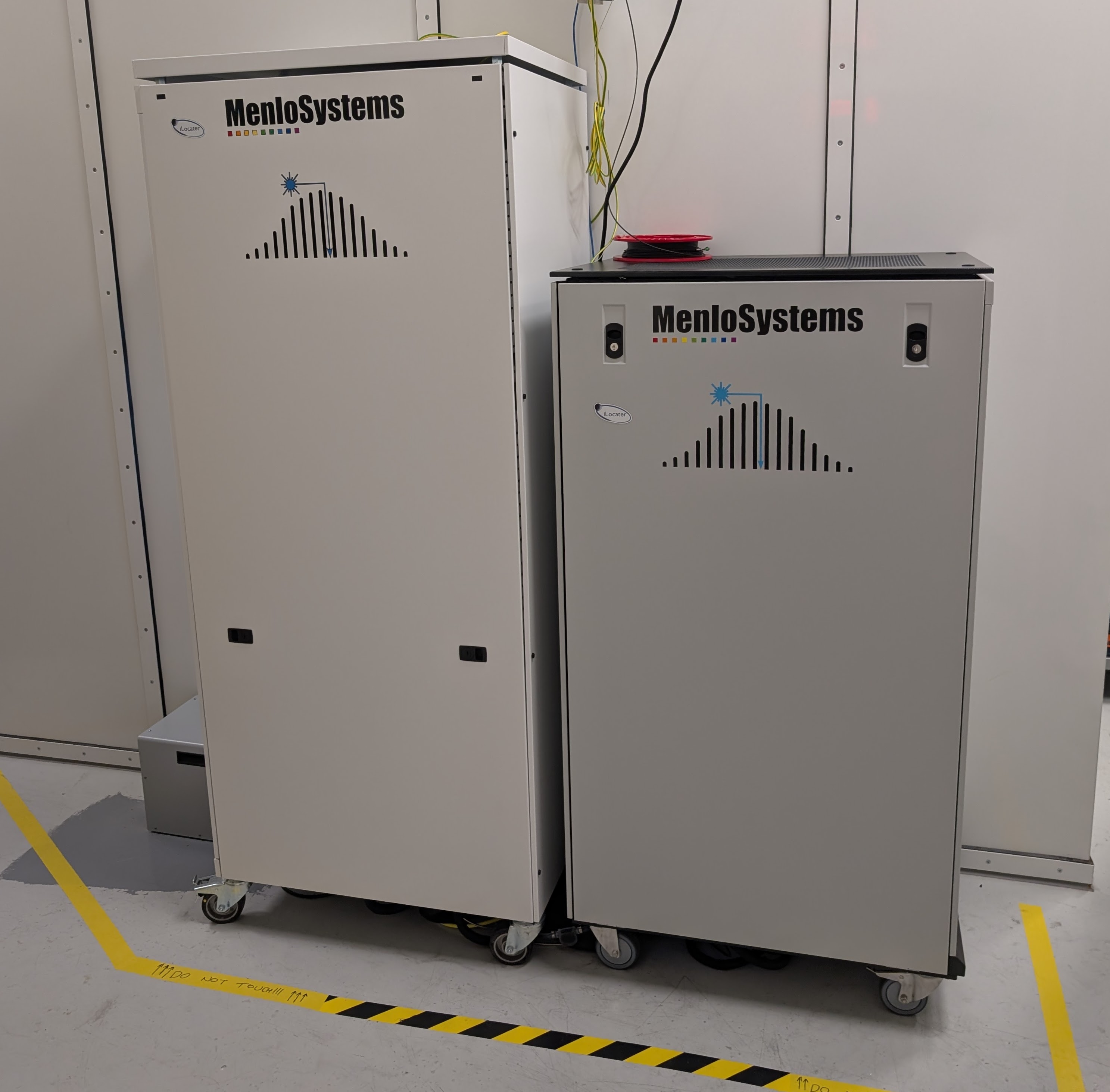}
        \caption{The LFC from Menlo Systems, which feeds the iLocater and PEPSI spectrographs at the LBT.}
        \label{fig:LFC}
    \end{minipage}
    
    \vspace{0.6cm} % Adds spacing between the figures

    \begin{minipage}{0.47\textwidth}
        \centering
        \includegraphics[width=\textwidth]{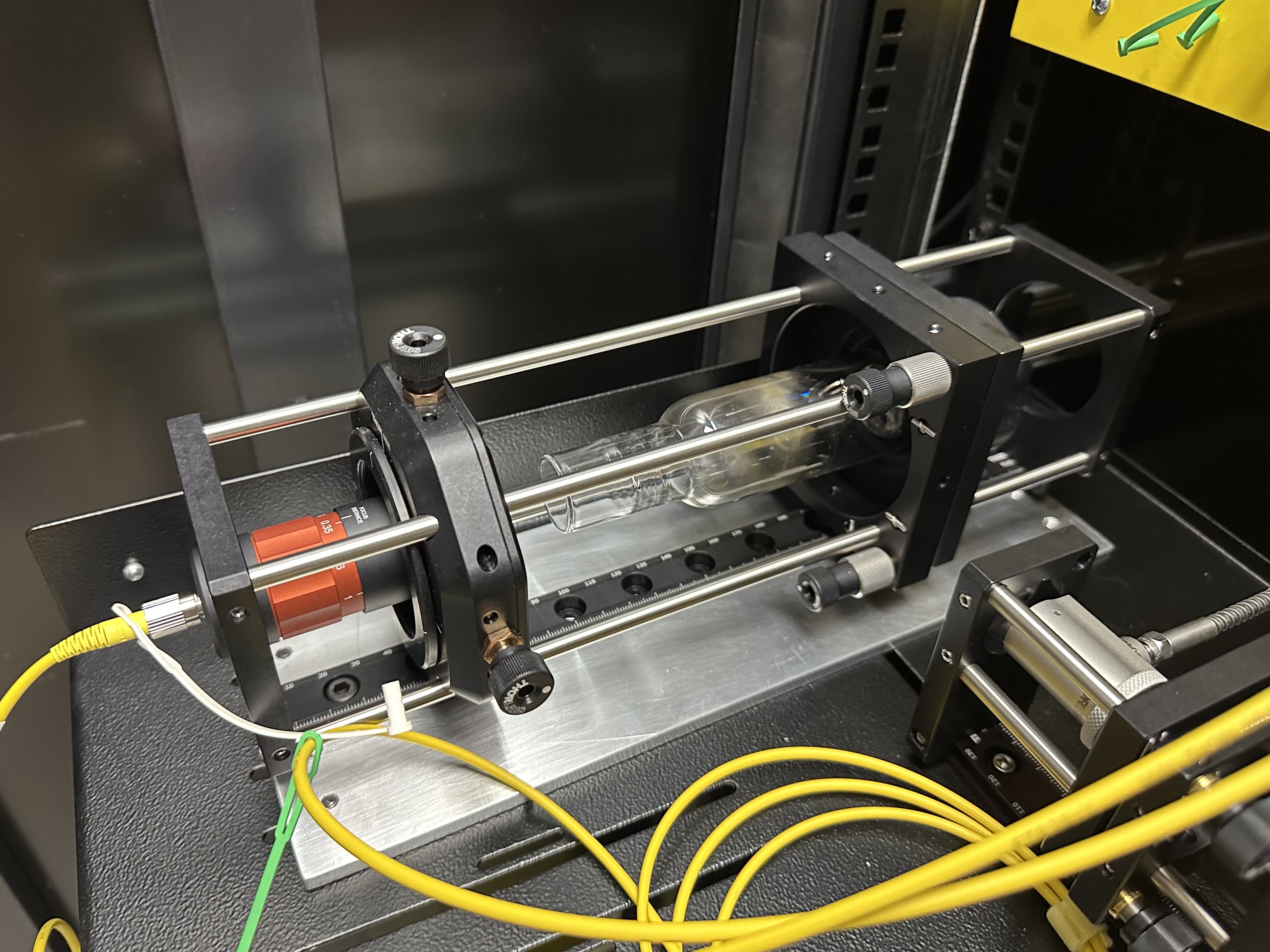}
        \caption{The iLocater UNe injection system.}
        \label{fig:UNe}
    \end{minipage}\hfill % Fills the space between them
    \begin{minipage}{0.47\textwidth}
    \centering
        \includegraphics[width=0.95\textwidth]{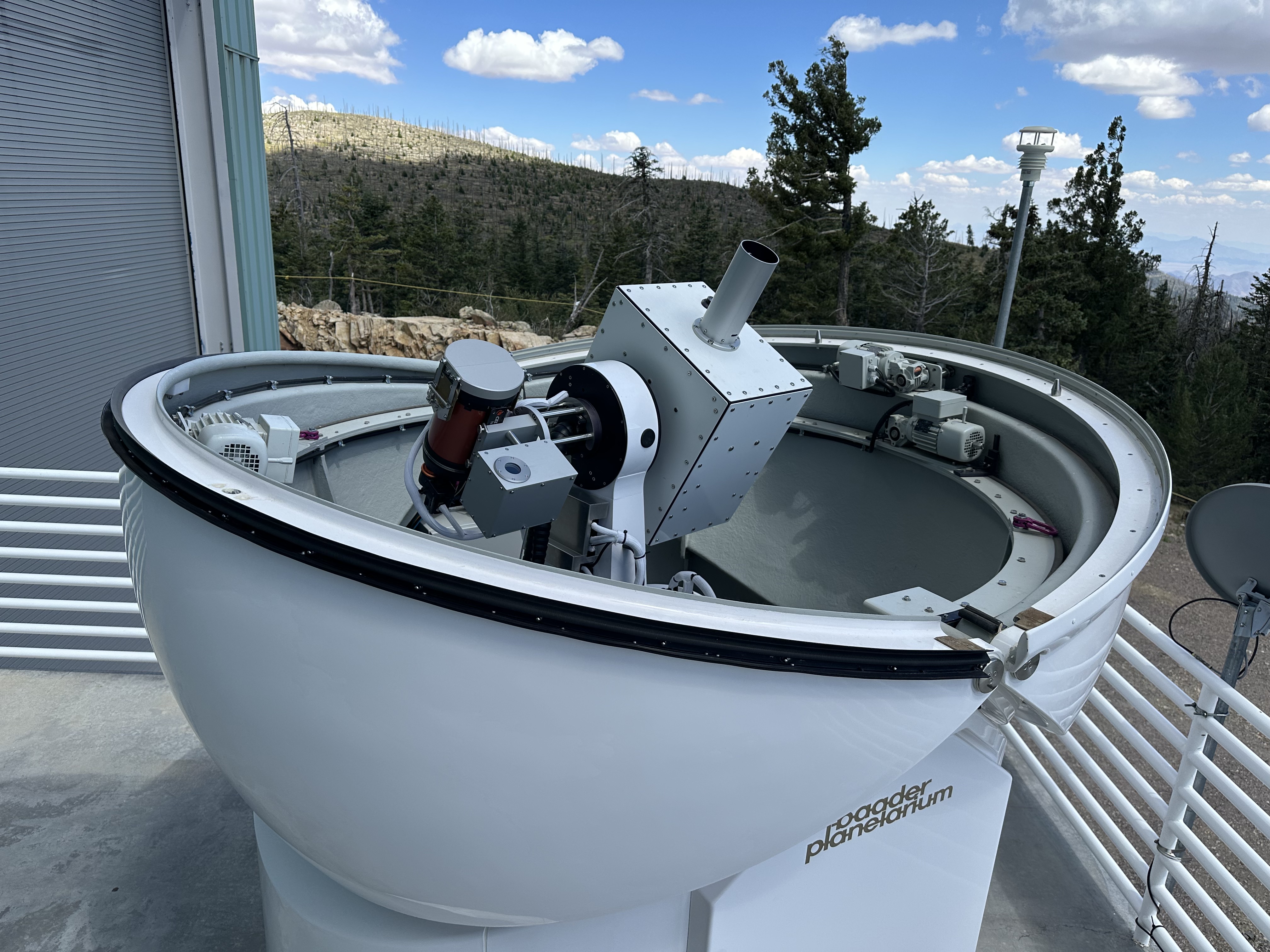}
        \caption{The iLocater solar feed mounted as part of the PEPSI solar telescope infrastructure.}
        \label{fig:solar}
    \end{minipage}

\end{figure}

%-------------------------------------------------------------------------
\subsection{Fabry-Pérot Etalon} \label{subsec:etalon}

A stabilized and optically monitored Fabry-Pérot etalon \cite{Gurevich2014, Kitzler2026} is used to generate a stable set of wavelength peaks for calibration and assess optical drifts within the instrument. This system can be illuminated using a halogen source (\S\ref{subsec:halogen_light_source}) or using a collimated SuperK COMPACT white light laser from NKT Photonics (Fig. \ref{fig:superk_compact_laser_and_etalon}). The collimated SuperK COMPACT beam passes through a custom notch filter from Optigrate to suppress the \qty{1064}{\nano\metre} peak present in the spectrum, and is then coupled into a SMF using a reflective collimator from Thorlabs (Part RC04FC-P01).

%-------------------------------------------------------------------------
\subsection{Halogen Light Source} \label{subsec:halogen_light_source}

A Thorlabs SLS201L halogen light source provides spectral order tracing and blaze function correction (Fig.~\ref{fig:halogen}). This is directly coupled into a SMF.

%-------------------------------------------------------------------------
\subsection{Laser Frequency Comb} \label{subsec:laser_frequency_comb}

A laser frequency comb (LFC) from Menlo Systems (Fig.~\ref{fig:LFC}) is used to calibrate wavelength features across the detector and track optical drifts. The system covers the entire instrument bandpass with a free-spectral-range of \qty{13}{\giga\hertz}. The LFC output is directly coupled into a SMF.

%-------------------------------------------------------------------------
\subsection{Uranium Neon Hollow Cathode Lamp} \label{subsec:uranium_neon_cathode_lamp}

A uranium neon (UNe) hollow cathode lamp from Photron (P863) provides absolute wavelength calibration (Fig.~\ref{fig:UNe}). This bulb is fixed in a cage mount system from Thorlabs and is coupled into a SMF using transmissive fiber collimating hardware (Part C40FC-C). The collimator and bulb have lateral and tip/tilt adjustment, respectively, to maximize input coupling.

%-------------------------------------------------------------------------
\subsection{Solar Feed} \label{subsec:solar_feed}

A solar feed provides disk-integrated sunlight to iLocater. The solar feed is comprised of an integrating sphere mounted onto an existing solar tracking infrastructure that feeds the PEPSI spectrograph (Fig.~\ref{fig:solar}). The use of an integrating sphere spatially scrambles the Sun's disk and couples this disk-integrated light into a \qty{200}{\micro\metre} fiber. This fiber then feeds into the iLocater room in the LBT. Here, the light is re-imaged using a pair of Thorlabs OAP fiber collimators (Parts RC04SMA-P01 and RC02FC-P01) and coupled into the SMF that feeds the instrument.

%=========================================================================
\section{Switching System Requirements and Schematic} \label{sec:switching_system_schematic}

The five calibration sources for iLocater must illuminate five distinct paths into the instrument. Two paths route directly to the spectrograph via fiber splitters on the SX and DX feeds (SX direct and DX direct). Two additional paths inject light into the calibration channels of the SX and DX fiber-coupling optical systems (SX acquisition camera and DX acquisition camera). The fifth path feeds the simultaneous calibration fiber (cryostat feed). Each of these illumination pathways must be independently controlled and deliver approximately equal intensity. Furthermore, the optical switching system must support the specific illumination modes required by the APERO data reduction pipeline \cite{Cook2022}, used by iLocater.

\begin{figure}[ht]
    \centering
    \includegraphics[width=\textwidth]{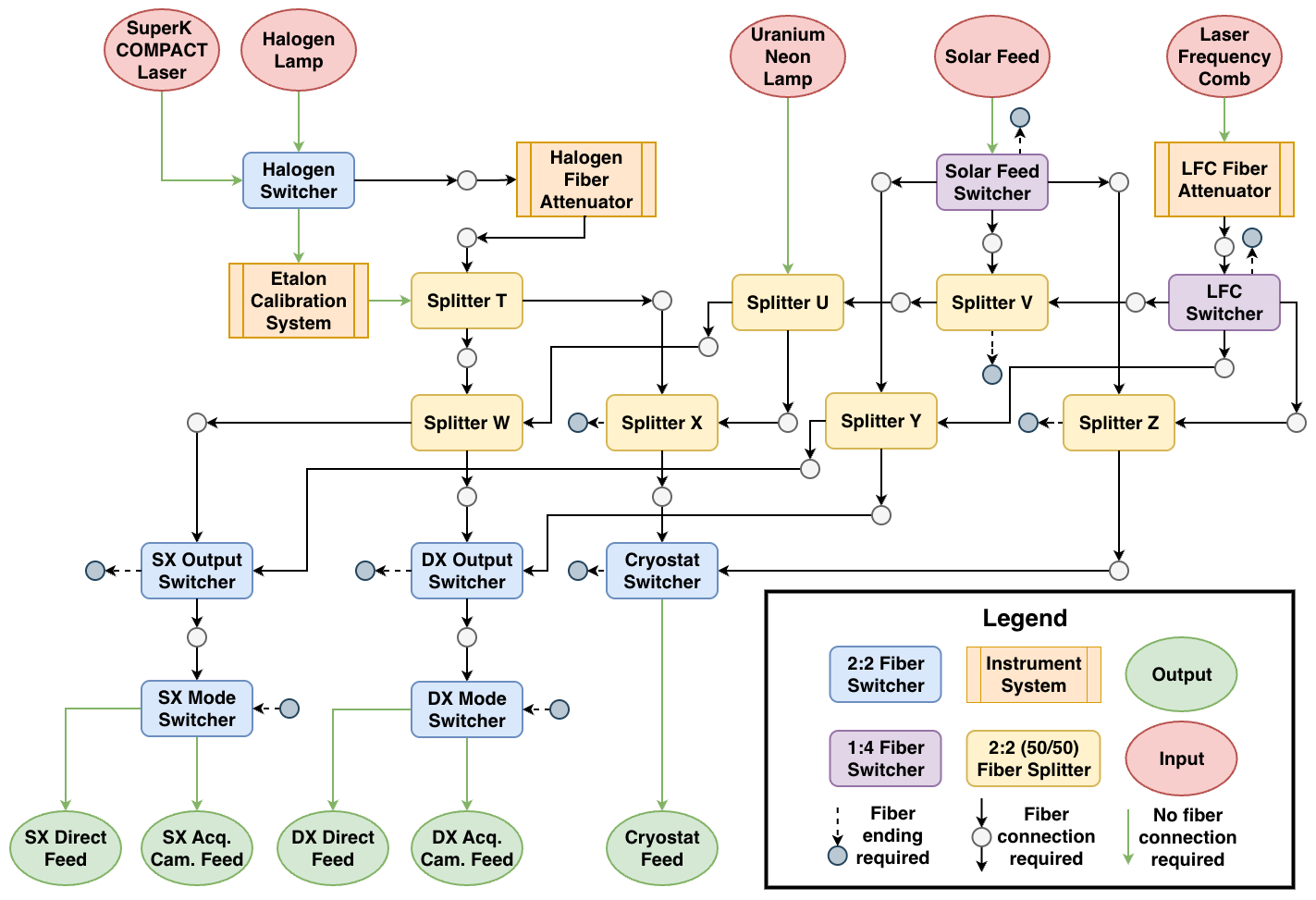}
    \caption{Schematic of the core components in the iLocater calibration system. Specialized switching and splitting hardware routes light from the illumination sources to the instrument's injection points. Each source can be independently directed to any output.}
    \label{fig:switching_system_schematic}
\end{figure}

As previously discussed, maintaining light strictly within SMFs minimizes optical losses and enhances the instrument's long-term operational stability. With the extensive use of SMFs in the telecommunications industry, commercial products exist for both switching single-mode fiber sources (i.e. fiber switchers) and mixing their signals (i.e. fiber splitters or mixers), presenting a potential solution for meeting iLocater's routing requirements. Following several trade studies to identify hardware compatible with the iLocater bandpass, we developed a system that integrates these commercial components. The resulting switching and splitting schematic is illustrated in Fig.~\ref{fig:switching_system_schematic}.

%=========================================================================
\section{System Hardware and Mechanical Design} \label{sec:mechanical_design_and_hardware}

\begin{figure}[ht]
    \centering
        \begin{minipage}{0.48\textwidth}
        \centering
        \includegraphics[width=0.85\textwidth]{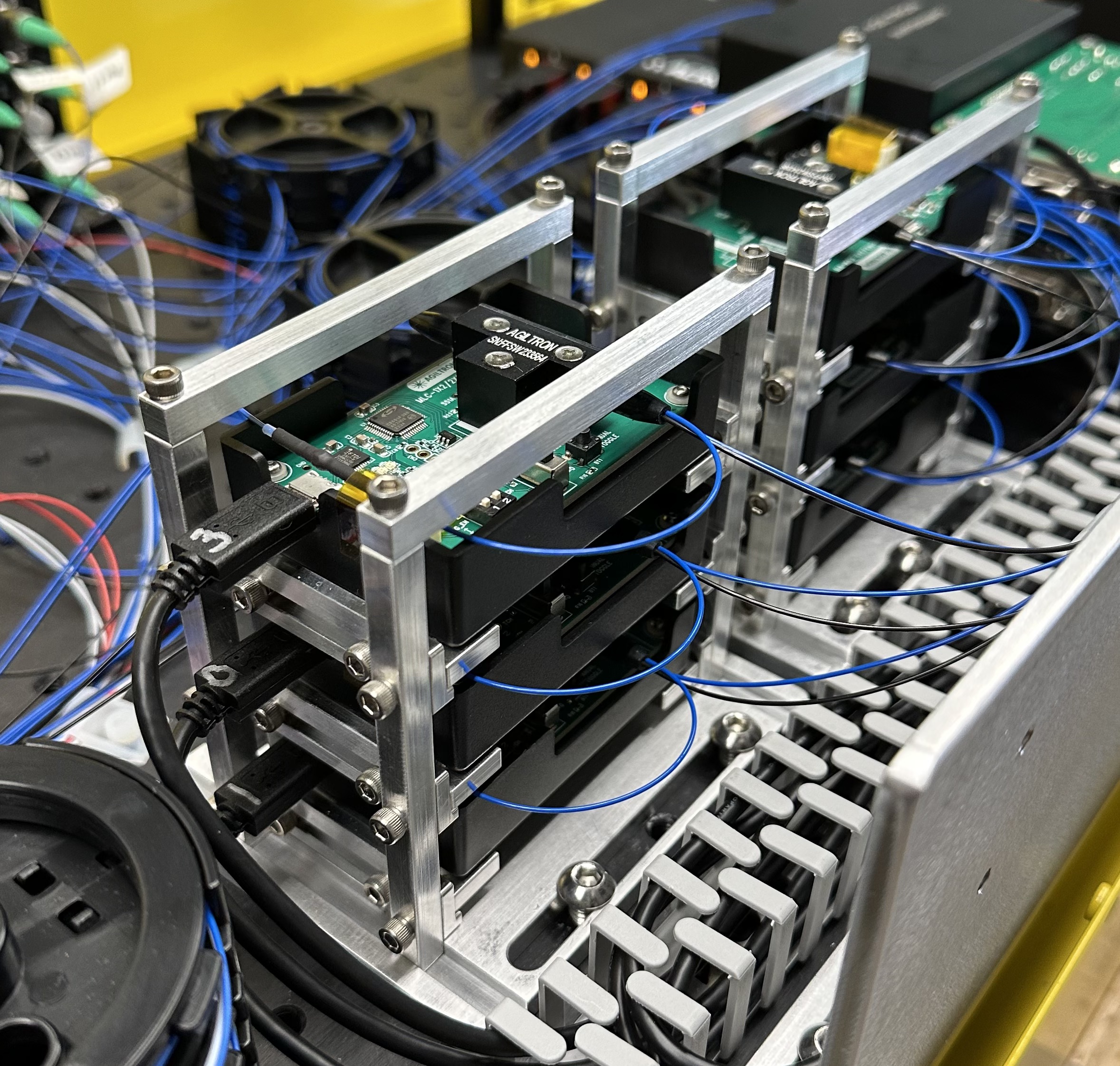}
        \caption{Two sets of three Agiltron 2$\times$2 Broadband Optical Switch Modules mounted in separate rack holders. Since these are electrical components and their internal moving micro-electro-mechanical systems (MEMS) parts have a finite lifespan, these units are designed to slide out of their racks to allow for easy replacement. They have built-in PCB drivers and a USB interface. The switches are configured for \qty{1060}{\nano\metre} wavelength with SM980 fiber, \qty{900}{\micro\metre} tubing, and FC/APC connectors. Part number: FFSM-22L109333.}
        \label{fig:2x2_switches}
    \end{minipage}\hfill % Fills the space between them
    \begin{minipage}{0.48\textwidth}
    \centering
        \includegraphics[width=0.85\textwidth]{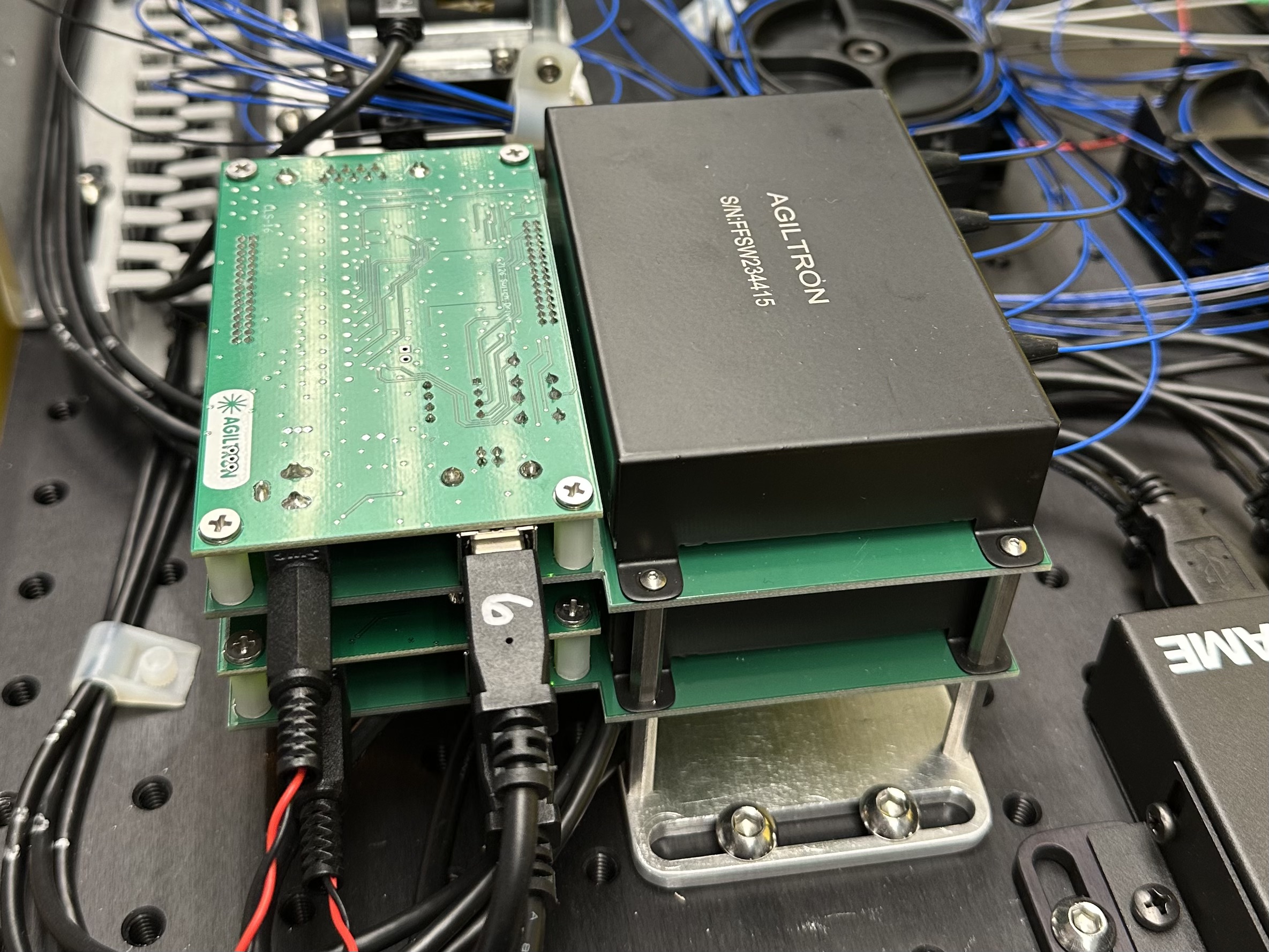}
        \caption{Two stacked Agiltron Fiber-Fiber\texttrademark{} 1$\times$4 Fiber Optical Switches. This stacked configuration minimizes the required footprint within the enclosure and facilitates cleaner cable management. They use MEMS components and have a built-in PCB driver with a USB interface. These switches are also configured for \qty{1060}{\nano\metre} wavelength with SM980 fiber, \qty{900}{\micro\metre} tubing, and FC/APC connectors. Part number: FFSW-14610933321.}
        \label{fig:1x4_switches}
    \end{minipage}

    \vspace{0.6cm} % Adds spacing between the figures
\end{figure}

\begin{figure}[p]
    \centering
    \begin{minipage}{0.47\textwidth}
        \centering
        \includegraphics[width=\textwidth]{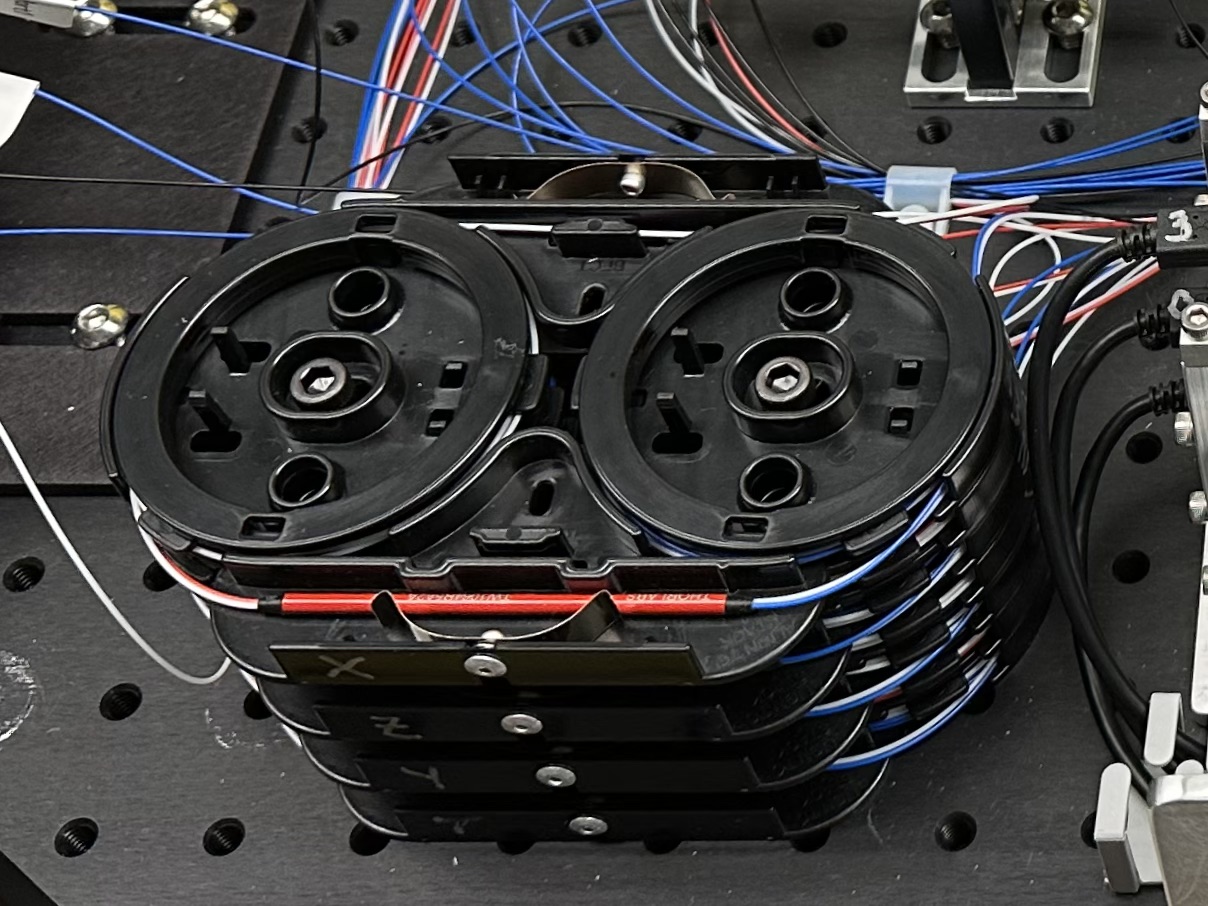}
        \caption{Seven Thorlabs TW1064R5A2A 2$\times$2 Wideband Fiber Optic Couplers stacked in Thorlabs BFCT Passive Component Fiber Trays, which secure the units and manage excess fiber length. These 50:50 splitters evenly mix light from the two inputs, distributing 50\% of the total intensity to each output. They are specified for \qty{1064}{\nano\metre} wavelength and are terminated with FC/APC fiber heads.}
        \label{fig:fiber_splitters}
    \end{minipage}\hfill % Fills the space between them
    \begin{minipage}{0.47\textwidth}
        \centering
        \includegraphics[width=\textwidth]{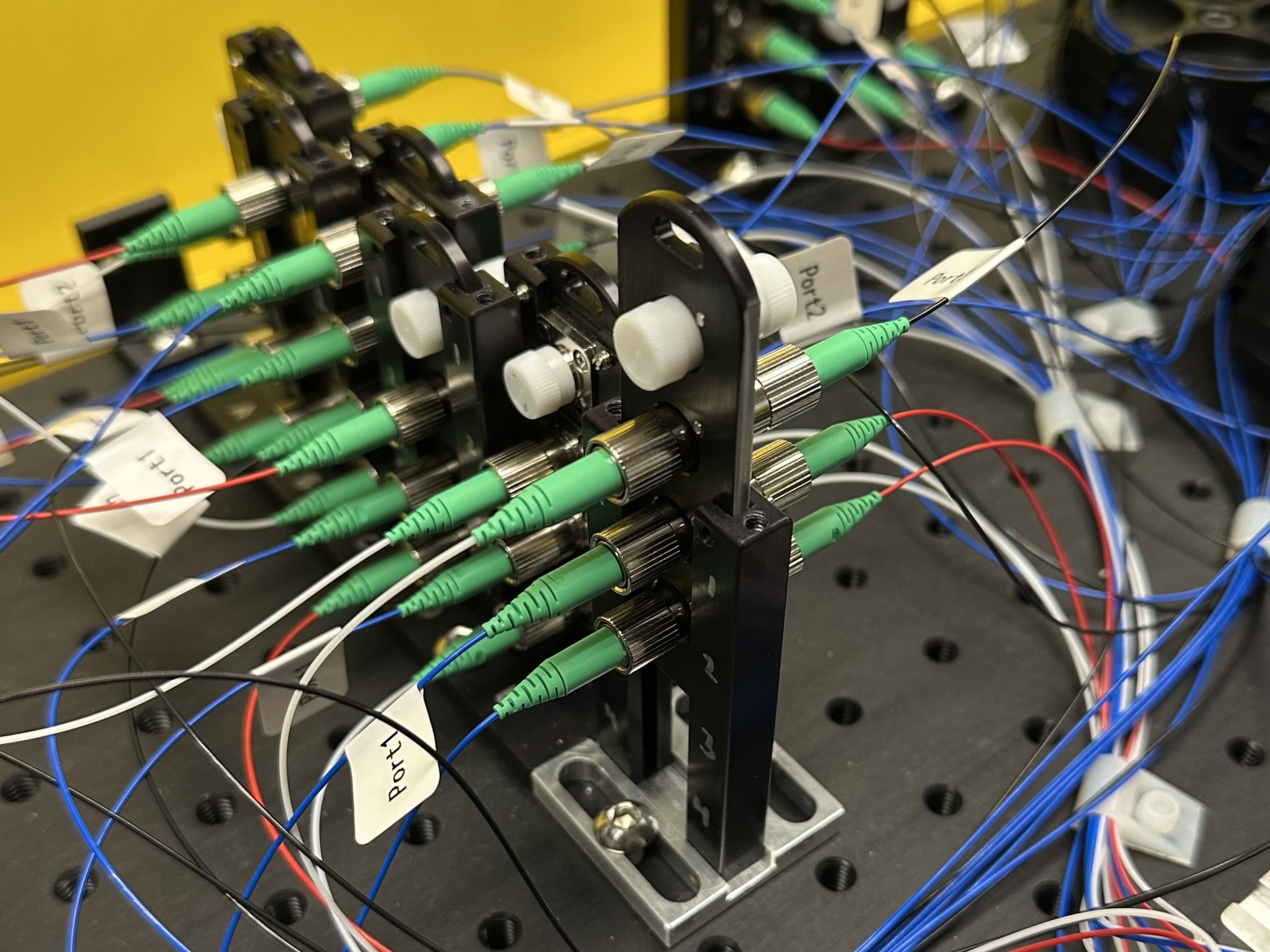}
        \caption{Six sets of four Thorlabs ADAFC3 FC/APC to FC/APC Mating Sleeves mounted on custom cassettes that can be lifted and locked in place by internal spring-ball plungers to access the bottom fiber heads. Each gray circle in the schematic in Fig.~\ref{fig:switching_system_schematic} requires a mating sleeve. Additional mating sleeves are used for one-sided fiber endings, shown by dotted lines going into dark blue circles in Fig.~\ref{fig:switching_system_schematic}.}
    \label{fig:fiber_connectors}
    \end{minipage}
    
    \vspace{0.6cm} % Adds spacing between the figures
     % Second Figure
    \begin{minipage}{0.48\textwidth}
        \centering
        \includegraphics[trim=0 50 0 150, clip, width=\textwidth]{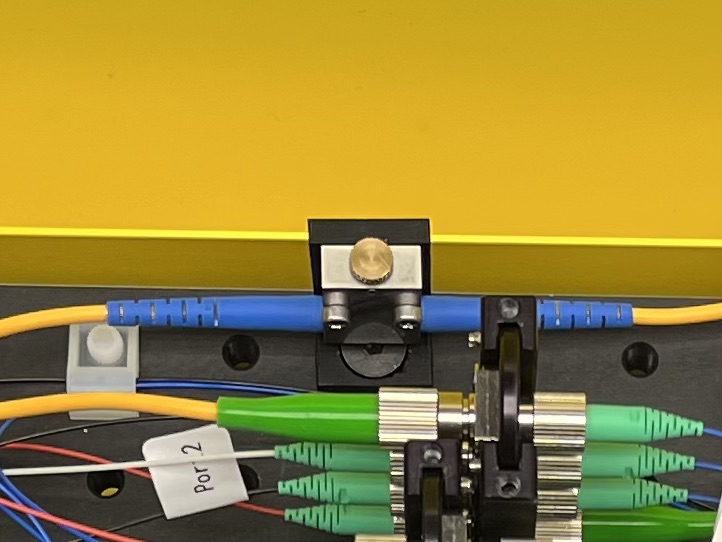}
        \caption{Two Thorlabs VOA1064-APC SM Variable Attenuators are used to dim the halogen (and consequently the SuperK COMPACT laser) and the laser frequency comb. The attenuation is manually adjusted with a knurled knob on top of the component and is configured to operate at \qty{1064}{\nano\metre} wavelengths. They have an attenuation range of 1.3 to 50 dB (0.001 to 74.1\% transmission range).}
        \label{fig:fiber_attenuator}
    \end{minipage}\hfill % Fills the space between them
    \begin{minipage}{0.48\textwidth}
    \centering
        \includegraphics[width=\textwidth]{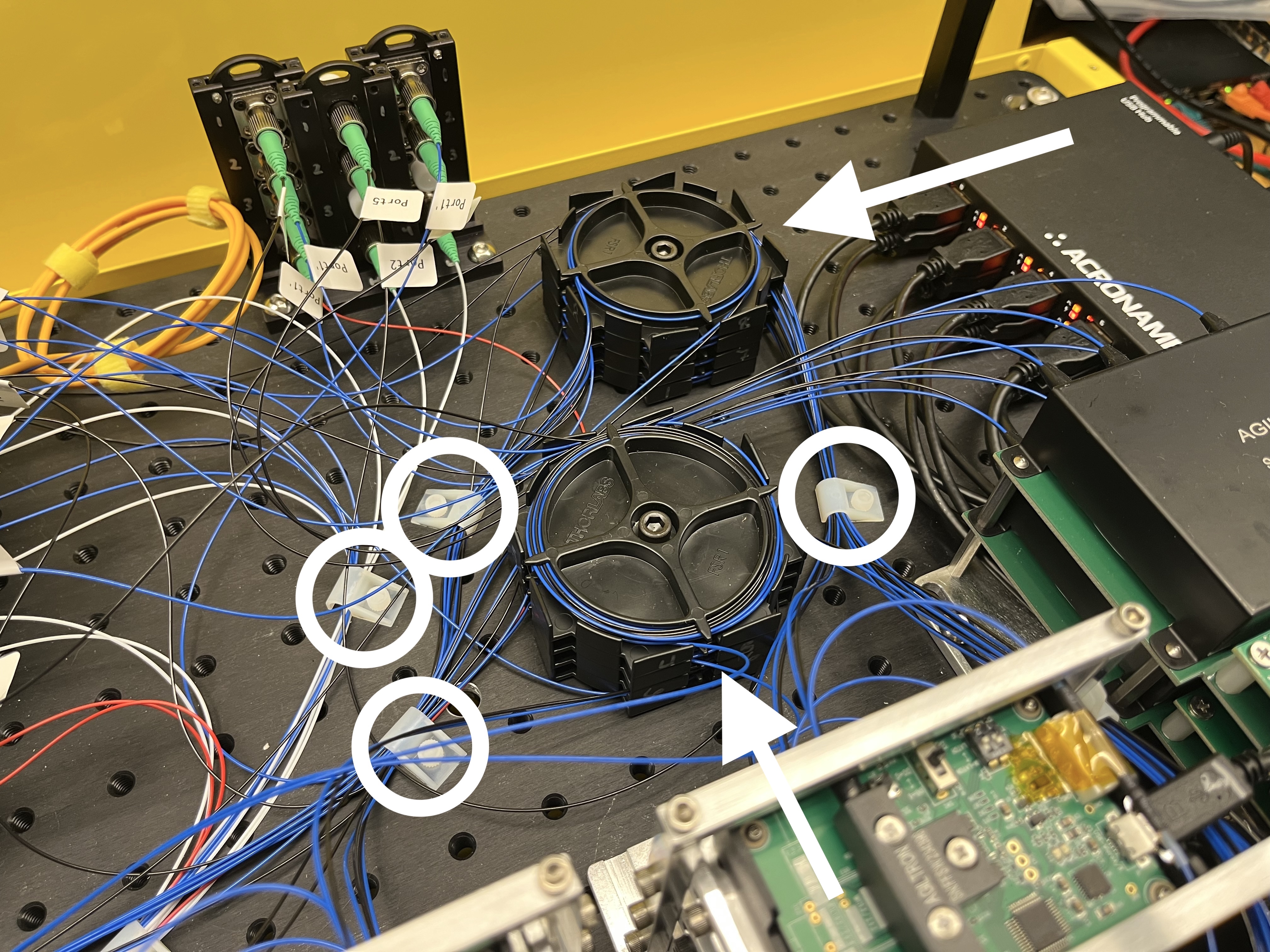}
        \caption{Two stacks of four Thorlabs FSR1 Storage Reels hold spools of excess fiber, shown with white arrows. McMaster-Carr 3192T42 Routing Clamps are used to tie down bundles of fiber to the breadboard at the base of the enclosure, shown in white circles. This prevents the fibers from shifting due to vibrations and forces during shipping. Adequate slack is given to the fiber connector cassettes for safe lifting.}
        \label{fig:cable_management}
    \end{minipage}

\end{figure}

The core hardware for fiber switching and mixing is sourced from Agiltron and Thorlabs, respectively. We utilize standard Agiltron 2$\times$2 (Part FFSM-22L109333) and 1$\times$4 (Part FFSW-14610933321) switchers, which were custom-fabricated using the same SM980 SMFs used throughout iLocater (Fig.~\ref{fig:2x2_switches} and Fig.~\ref{fig:1x4_switches}). For 50:50 signal mixing across both outputs, we employ standard Thorlabs 2$\times$2 fiber splitters (Part TW1064R5A2A) (Fig.~\ref{fig:fiber_splitters}). All fiber components are terminated with FC/APC connectors and joined using Thorlabs fiber-to-fiber mating sleeves (Part ADAFC3) (Fig.~\ref{fig:fiber_connectors}). To manage brighter calibration inputs, specifically the halogen and LFC sources, Thorlabs inline fiber attenuators (Part VOA1064-APC) reduce intensities to appropriate instrument levels (Fig.~\ref{fig:fiber_attenuator}). Finally, because each component includes approximately \qty{1}{\metre} of attached fiber, P-clips and fiber spool holders are integrated to secure the cabling during shipping and assembly (Fig.~\ref{fig:cable_management}).

\begin{figure}[ht]
    \centering
    \begin{tabular}{c c} %% tabular useful for creating an array of images 
        \includegraphics[trim=30 0 10 0, clip, width=0.46\textwidth]{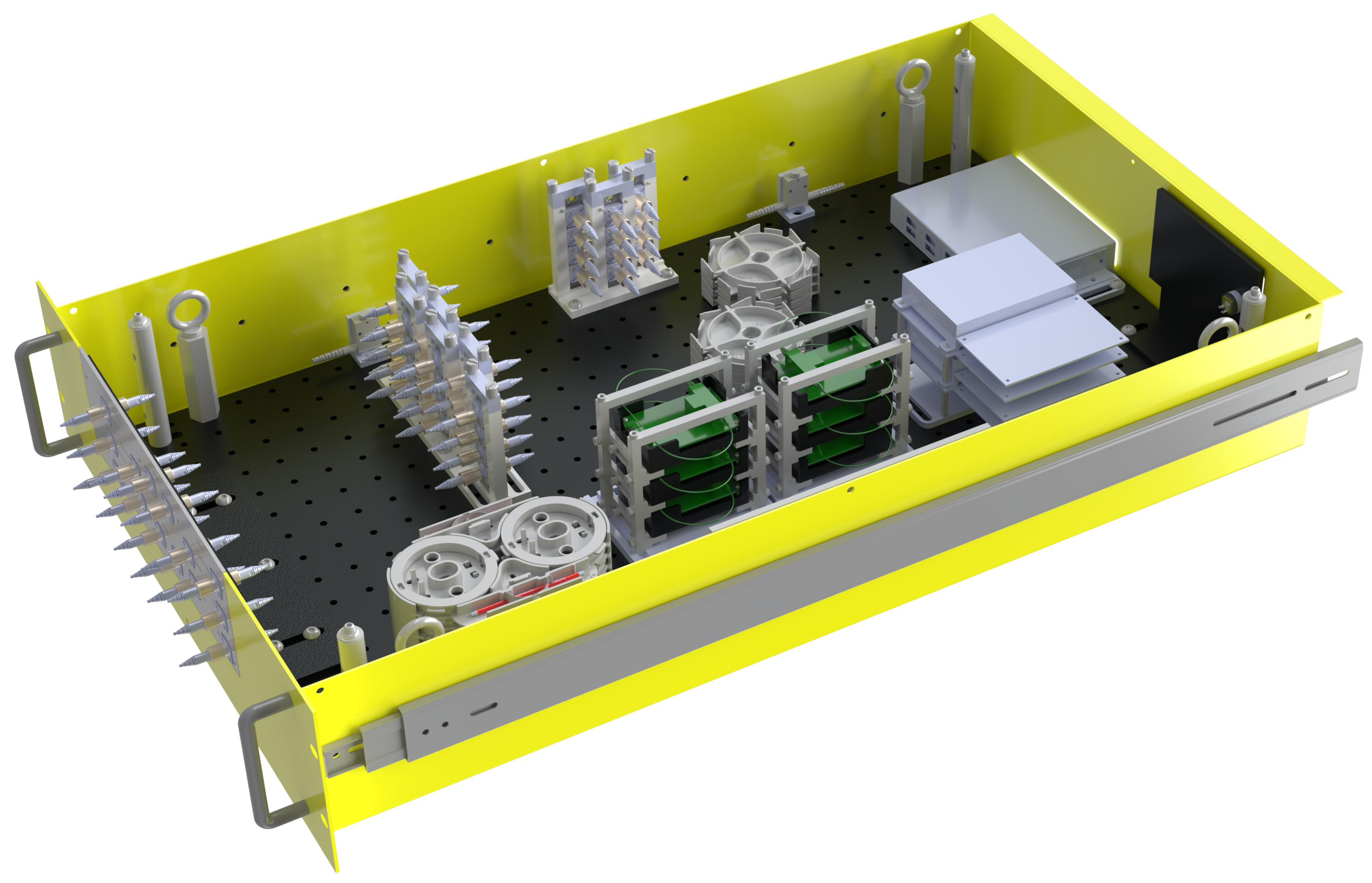} & \includegraphics[width=0.46\textwidth]{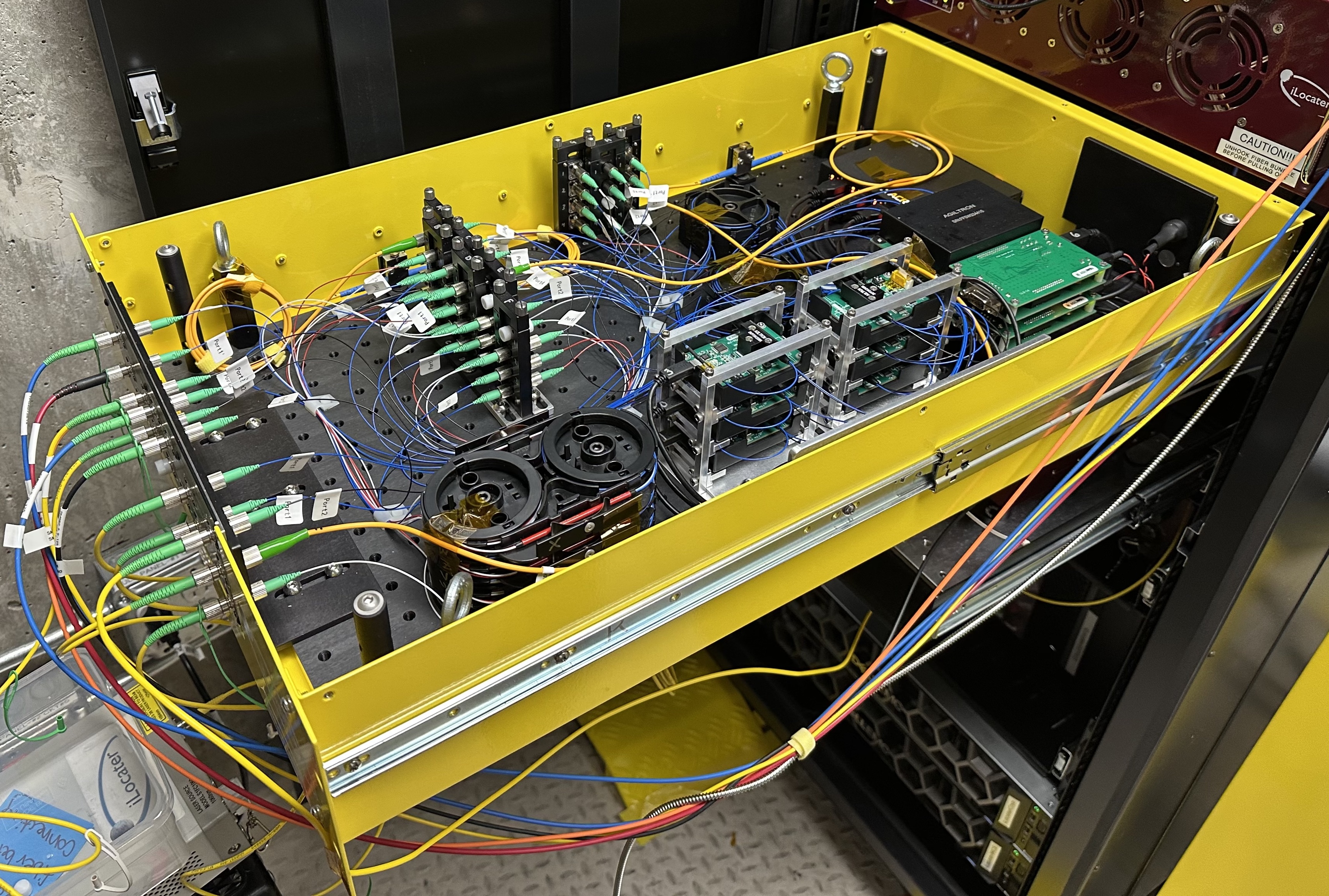}
    \end{tabular}
    \caption{Left: A CAD rendering of the switcher box in its stowed/operational configuration. Right: The rack-mounted fiber switcher enclosure, shown fully extended on its rails with the top cover removed.}
    \label{fig:box_cad_rendering_picture}
    \vspace{0.6cm} % Adds spacing between the figures
    \begin{minipage}{0.47\textwidth}
        \centering
        \includegraphics[width=\textwidth]{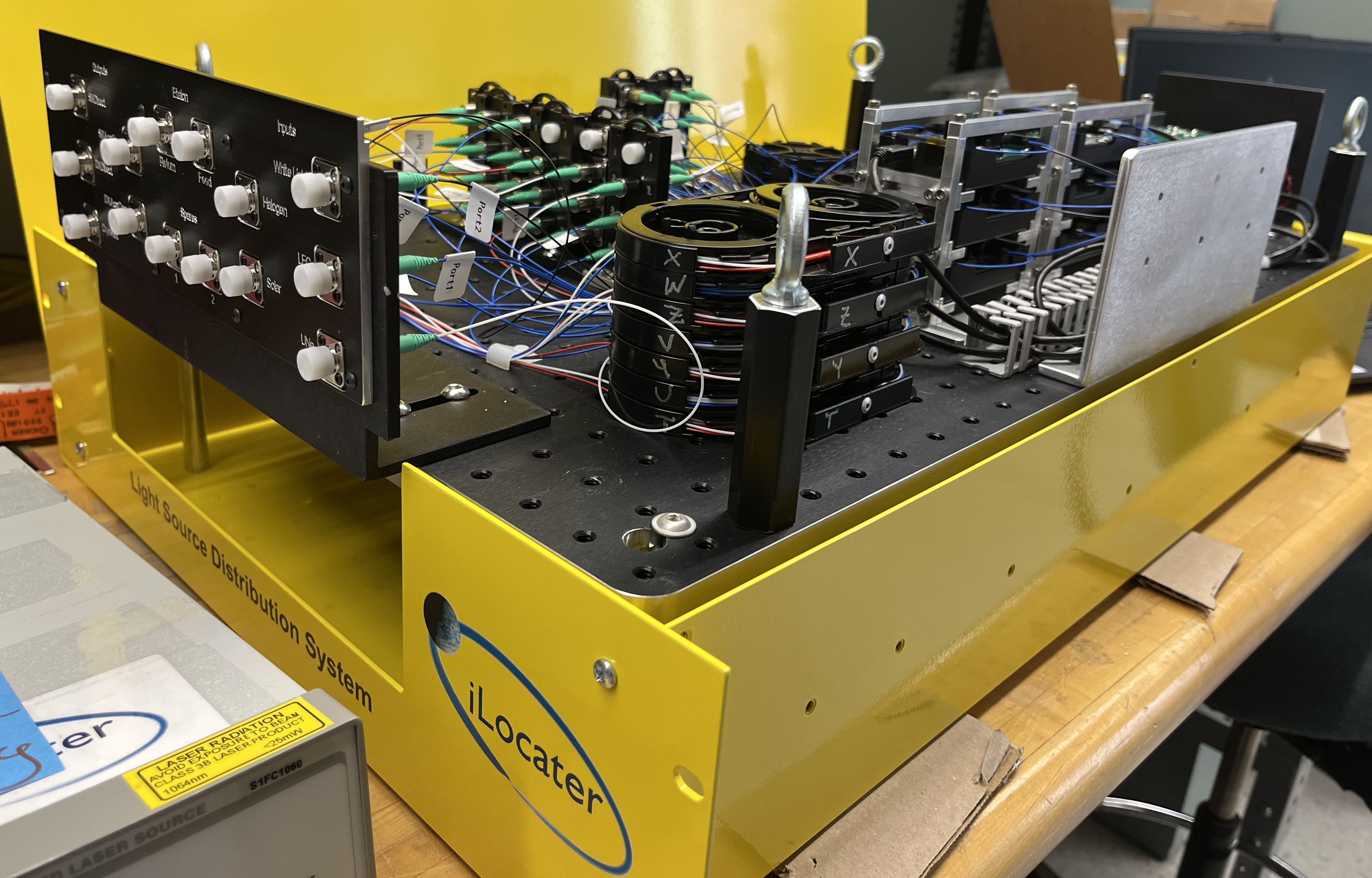}
        \caption{Lifted configuration of the fiber switcher box. Using four eyebolts located on the corners of the breadboard, the base can be lifted out of the enclosure and locked in place onto four posts for maintenance and accessibility.}
        \label{fig:lifted_configuration}
    \end{minipage}\hfill % Fills the space between them
    \begin{minipage}{0.47\textwidth}
        \centering
        \includegraphics[width=\textwidth]{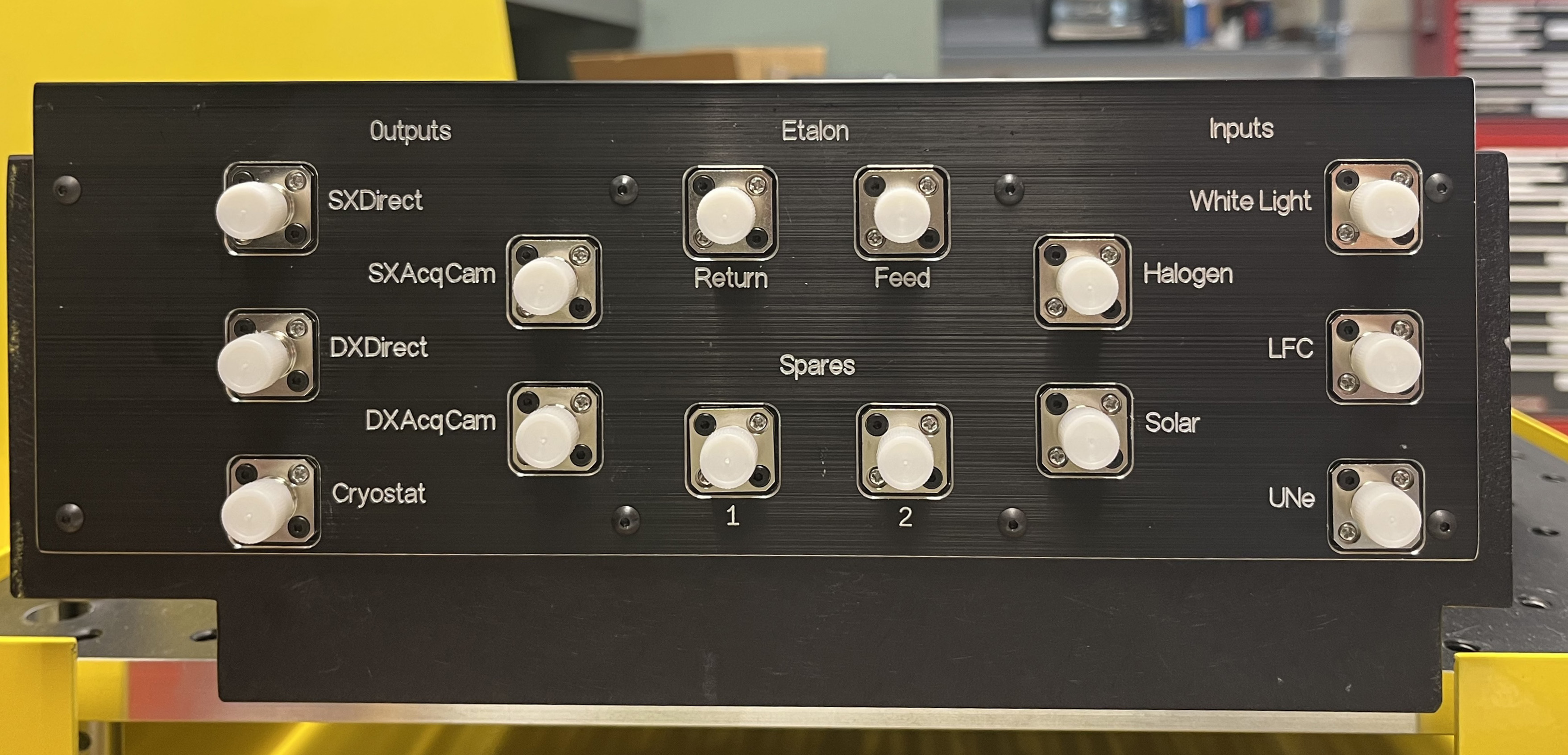}
        \caption{An anodized aluminum front plate with engraved labels for inputs and outputs of the box. Cutouts in the front plate allow fiber mating sleeves to sit flush to the front of the box and cutouts in the front of the enclosure allow the plate to sit flush and prevent contaminants, like dust, from entering the box.}
        \label{fig:front_plate}
    \end{minipage}
\end{figure}

To ensure easy maintenance and compatibility with space limitations, the entire switching system is integrated into a 3U rack-mounted enclosure. It features a removable lid, modular fiber connection cassettes, switcher racks, and a removable breadboard panel for quick component access. The enclosure is mounted on two rails within a server rack, along with the other calibration light sources. Figure~\ref{fig:box_cad_rendering_picture} shows the final computer-aided design (CAD) rendering and active configuration of the as-built system.

The enclosure houses all major hardware components: six 2$\times$2 switches, two 1$\times$4 switches, seven 2×2 50:50 splitters, fiber-to-fiber mating sleeves, two fiber attenuators, and a USB hub (Acroname USBHub3+). The fiber switches are active components that require power and control which is managed with USB control via the instrument control system. Splitters, connectors, and attenuators are entirely passive components.

For easy maintenance over time and accessibility, the internal breadboard, fastened to the base of the enclosure, can be removed using lifting straps and placed onto four posts within the enclosure. The lifted configuration is shown in Fig.~\ref{fig:lifted_configuration}. A front plate is mounted to the internal breadboard and is slotted flush with the front of the enclosure. Engravings on the front plate label each of the inputs and outputs of the switching system, along with spares for future use. This plate can be seen in Fig.~\ref{fig:front_plate}. The front plate fiber connections do not need to be removed when transitioning from the stowed to lifted configurations, however, the power and control cables to the USB hub must be disconnected.

%=========================================================================
\section{Preliminary Laboratory Testing Results}
\label{sec:preliminary_laboratory_testing_results}

End-to-end throughput is a key metric for evaluating system performance. We assessed the as-built system by injecting a benchtop laser of known power into each input port and recording the resulting power at each output. Tables \ref{tab:testing_results_left} through \ref{tab:testing_results_LFC} present these results, comparing them against two baselines: an idealized theoretical maximum (assuming perfect 50:50 power division by the splitters and zero insertion loss) and a more realistic expected throughput based on component specifications. For reference, these tables also list the number of fiber-fiber connections, 2$\times$2 switches, and 1$\times$4 switches in each pathway. In Tables \ref{tab:testing_results_solar} and \ref{tab:testing_results_LFC}, ``individual illumination" denotes routing light separately to either the SX/DX or the cryostat feeds (via Splitters Y or Z in Fig.~\ref{fig:switching_system_schematic}, respectively). Conversely, ``simultaneous illumination" refers to feeding both pathways concurrently using Splitter V.

Several factors cause losses in throughput across the fiber switching system. With the size of the SMF cores (\qty{6}{\micro\metre}), the concentricity of the fiber within its ferrule and the alignment of fiber connectors within their mating sleeves play a large role in potential insertion losses of many percent at each connector. From vendor specifications, the expected maximum losses of the Agiltron 2$\times$2 and 1$\times$4 switches are about 0.48 dB (10.5\%) and 1.22 dB (25\%), respectively, including insertion loss (0.4/1.0 dB), wavelength dependent loss (0.01/0.1 dB), polarization dependent loss (0.05/0.1 dB), and repeatability loss ($\pm$0.02 dB) \cite{Agiltron2x2, Agiltron1x4}. Thorlabs splitters have a typical insertion loss of less than 0.7 dB, with secondary wavelength-dependent effects that can impact throughput \cite{ThorlabsSplitter}. Thorlabs mating sleeves have a typical insertion loss of less than 0.5 dB \cite{ThorlabsADAFC3}. Dirty fiber heads also cause losses, but this is mitigated by thoroughly cleaning and checking the state of the fiber heads before insertion into mating sleeves.

Overall, the measured system throughput aligns with the expected performance when accounting for the typical and worst-case hardware losses outlined in vendor specifications. The primary driver of signal reduction is SMF insertion loss. To manage this, components identified during testing as having high insertion losses were strategically paired with bright calibration sources (e.g., the solar and LFC feeds) where such losses are easily tolerated. While replacing physical fiber-to-fiber connections with fusion splicing could further boost throughput, current performance levels make this optimization unnecessary.

\begin{table}[htb]
  \centering
        \caption{Throughput for the UNe lamp, SuperK COMPACT laser, and halogen light source. A: SX direct, B: SX acquisition camera, C: DX direct, D: DX acquisition camera, E: Cryostat.}
    \resizebox{\textwidth}{!}{% 
    \small
    \begin{tabular}{|l!{\vrule width 1.5pt}c|c|c|c|c!{\vrule width 1.5pt}c|c|c|c|c!{\vrule width 1.5pt}c|c|c|c|c|}
        \hline
        \rule[-1ex]{0pt}{3.5ex} Source & 
        \multicolumn{5}{c!{\vrule width 1.5pt}}{UNe} & 
        \multicolumn{5}{c!{\vrule width 1.5pt}}{SuperK COMPACT} & 
        \multicolumn{5}{c|}{Halogen} \\
        \hline
        \rule[-1ex]{0pt}{3.5ex} Output & A & B & C & D & E & A & B & C & D & E & A & B & C & D & E \\
        \Xhline{2pt}
        \rule[-1ex]{0pt}{3.5ex} Number of Fiber Connections & 3 & 3 & 3 & 3 & 2 & 3 & 3 & 3 & 3 & 2 & 5 & 5 & 5 & 5 & 4 \\
        % \hline
        \rule[-1ex]{0pt}{3.5ex} Number of 2x2 Switches & 2 & 2 & 2 & 2 & 1 & 3 & 3 & 3 & 3 & 2 & 3 & 3 & 3 & 3 & 2 \\
        % \hline
        \rule[-1ex]{0pt}{3.5ex} Number of 1x4 Switches & 0 & 0 & 0 & 0 & 0 & 0 & 0 & 0 & 0 & 0 & 0 & 0 & 0 & 0 & 0 \\
        % \hline
        \rule[-1ex]{0pt}{3.5ex} Theoretical Max. Throughput [\%] & 25 & 25 & 25 & 25 & 25 & 25 & 25 & 25 & 25 & 25 & 25 & 25 & 25 & 25 & 25 \\
        % \hline
        \rule[-1ex]{0pt}{3.5ex} Predicted Throughput (Typ.) [\%] & 14.1 & 14.1 & 14.1 & 14.1 & 14.9 & 13.2 & 13.2 & 13.2 & 13.2 & 14.1 & 13.2 & 13.2 & 13.2 & 13.2 & 14.1 \\
        \Xhline{1.5pt}
        \rule[-1ex]{0pt}{3.5ex} Measured Throughput [\%] & 11.7 & 12.0 & 12.3 & 11.0 & 15.7 & 11.7 & 11.7 & 11.7 & 10.0 & 13.0 & 11.0 & 11.7 & 11.7 & 10.3 & 13.3 \\
        \hline
    \end{tabular}%
    }
    \label{tab:testing_results_left}

    \vspace{3ex}
        \caption{Throughput for the solar feed through two paths. A: SX direct, B: SX acquisition camera, C: DX direct, D: DX acquisition camera, E: Cryostat.}
        \centering
        \resizebox{0.8\textwidth}{!}{% 
        \small
        \begin{tabular}{|l!{\vrule width 1.5pt}c|c|c|c|c!{\vrule width 1.5pt}c|c|c|c|c|}
            \hline
            \rule[-1ex]{0pt}{3.5ex} Source & 
            \multicolumn{5}{c!{\vrule width 1.5pt}}{Solar (individual illumination)} & 
            \multicolumn{5}{c|}{Solar (simultaneous illumination)} \\
            \hline
            \rule[-1ex]{0pt}{3.5ex} Output & A & B & C & D & E & A & B & C & D & E \\
            \Xhline{2pt}
            \rule[-1ex]{0pt}{3.5ex} Number of Fiber Connections & 3 & 3 & 3 & 3 & 2 & 4 & 4 & 4 & 4 & 3 \\
            % \hline
            \rule[-1ex]{0pt}{3.5ex} Number of 2x2 Switches & 2 & 2 & 2 & 2 & 1 & 2 & 2 & 2 & 2 & 1 \\
            % \hline
            \rule[-1ex]{0pt}{3.5ex} Number of 1x4 Switches & 1 & 1 & 1 & 1 & 1 & 1 & 1 & 1 & 1 & 1 \\
            % \hline
            \rule[-1ex]{0pt}{3.5ex} Theoretical Max. Throughput [\%] & 50 & 50 & 50 & 50 & 50 & 12.5 & 12.5 & 12.5 & 12.5 & 12.5 \\
            % \hline
            \rule[-1ex]{0pt}{3.5ex} Predicted Throughput (Typ.) [\%] & 29.7 & 29.7 & 29.7 & 29.7 & 31.6 & 4.71 & 4.71 & 4.71 & 4.71 & 5.00 \\
            \Xhline{1.5pt}
            \rule[-1ex]{0pt}{3.5ex} Measured Throughput [\%] & 16.7 & 17.7 & 16.3 & 15.0 & 21.3 & 4.00 & 4.00 & 4.67 & 3.00 & 6.33 \\
            \hline
        \end{tabular}%
        }
        \label{tab:testing_results_solar}

    \vspace{3ex}
    \caption{Throughput for the LFC feed for two paths. A: SX direct, B: SX acquisition camera, C: DX direct, D: DX acquisition camera, E: Cryostat.}
    \centering
    \resizebox{0.8\textwidth}{!}{% 
    \small
    \begin{tabular}{|l!{\vrule width 1.5pt}c|c|c|c|c!{\vrule width 1.5pt}c|c|c|c|c|}
        \hline
        \rule[-1ex]{0pt}{3.5ex} Source & 
        \multicolumn{5}{c!{\vrule width 1.5pt}}{LFC (individual illumination)} & 
        \multicolumn{5}{c|}{LFC (simultaneous illumination)} \\
        \hline
        \rule[-1ex]{0pt}{3.5ex} Output & A & B & C & D & E & A & B & C & D & E \\
        \Xhline{2pt}
        \rule[-1ex]{0pt}{3.5ex} Number of Fiber Connections & 3 & 3 & 3 & 3 & 2 & 5 & 5 & 5 & 5 & 4 \\
        % \hline
        \rule[-1ex]{0pt}{3.5ex} Number of 2x2 Switches & 2 & 2 & 2 & 2 & 1 & 2 & 2 & 2 & 2 & 1 \\
        % \hline
        \rule[-1ex]{0pt}{3.5ex} Number of 1x4 Switches & 1 & 1 & 1 & 1 & 1 & 1 & 1 & 1 & 1 & 1 \\
        % \hline
        \rule[-1ex]{0pt}{3.5ex} Theoretical Max. Throughput [\%] & 50 & 50 & 50 & 50 & 50 & 12.5 & 12.5 & 12.5 & 12.5 & 12.5 \\
        % \hline
        \rule[-1ex]{0pt}{3.5ex} Predicted Throughput (Typ.) [\%] & 29.7 & 29.7 & 29.7 & 29.7 & 31.6 & 4.71 & 4.71 & 4.71 & 4.71 & 5.00 \\
        \Xhline{1.5pt}
        \rule[-1ex]{0pt}{3.5ex} Measured Throughput [\%] & 20.3 & 18.7 & 21.3 & 18.0 & 23.7 & 5.00 & 5.00 & 5.00 & 4.67 & 7.33 \\
        \hline
    \end{tabular}%
    }
    \label{tab:testing_results_LFC}
\end{table}

%=========================================================================
\section{Conclusions and Future Work}
\label{sec:conclusion_and_future_work}

Calibrating the iLocater spectrograph requires a comprehensive suite of light sources to illuminate key locations across the instrument. Because iLocater relies entirely on SMFs for spectrograph injection, we have developed a specialized system to couple our five distinct sources directly into SMFs. By leveraging commercial fiber switchers and splitters, this system distributes light throughout the instrument while providing precise, independent control over each input. The complete switching hardware, housed in a rack-mounted enclosure, is now fully installed and operational at the LBT. It is currently executing daily instrument calibrations, which will seamlessly provide the data necessary to characterize the system's long-term performance.

%=========================================================================
\acknowledgments
 
This material is based upon work supported by the National Science Foundation under Grant No. 1654125, 2108603, and 2408424, the National Aeronautics and Space Administration under Space Act Agreement No. 38232 and Grant No. 80NSSC24K1444, and the Mt. Cuba Astronomical Foundation. J. Crepp acknowledges partial support for the iLocater project from the Wolfe family and Potenziani family.

%=========================================================================
% References
\bibliography{report} % bibliography data in report.bib
\bibliographystyle{spiebib} % makes bibtex use spiebib.bst

\end{document}